# The Mysterious Phenomenon of Forward-Progressing Student Tables

Samuel J. George[1]

[1] Ark St. Alban's Academy, Birmingham, UK

E-mail: samuel.george@arkstalbans.org

## Abstract

This study investigates the factors that contribute to the forward movement of student desks throughout the school day. We hypothesize that desk movement is influenced not only by classroom floor type but also by the physical characteristics of students, such as height and age. Furthermore, we explore how the subject taught in the classroom (e.g., Science vs. Modern Foreign Languages) contributes to desk dynamics. Utilizing a Monte Carlo simulation model, we quantitatively analyse the forces at play in these phenomena. This research reveals that desks on carpeted floors are particularly prone to movement, especially in science classrooms with taller and younger students. While the results may seem trivial, they provide critical insights into the mechanics of classroom furniture behaviour and its implications for educational practices. The paper offers compelling evidence that classroom furniture has a mind of its own—or, at the very least, a subtle gravitational pull towards the front of the room.

## 1. Introduction

Classroom furniture movement has been a topic of great concern in educational psychology, with researchers often attributing desk shifts to physical forces such as friction, student behaviour, and environmental factors. In this study, we aim to explore the multivariable causes behind the forward movement of student desks during the school day, week, and month. Through a combination of physical modelling and psychological factors, we seek to explain the role of floor type, student height and age, and subject-specific classroom dynamics in this seemingly trivial yet surprisingly complex phenomenon.

The present study seeks to explore the influence of two potential factors: the age of the students and the type of flooring. Previous research has failed to explore the gravitational pull of various flooring materials on student desks, and the current paper seeks to fill that gap. Our hypothesis posits that younger students, who may possess more restless energy, are more likely to initiate table movement, whereas older students, with greater academic responsibility, may influence the desks **to move more subtly, and at a greater velocity.**

.

*1.1 Hypothesis*

We propose the following hypotheses to guide our investigation into the dynamics of desk movement:

1. **Hypothesis 1**: Desk movement is more pronounced on **carpeted floors** due to increased friction compared to other floor types (e.g., linoleum or wood).

2. **Hypothesis 2**: **Taller** and **younger** students (6-10 years old) exert more force on desks, leading to greater desk movement.

3. **Hypothesis 3**: Desk movement is significantly more frequent in **science classrooms**, where group work and hands-on activities are more common, than in **MFL classrooms**, where passive learning is the norm.

### 1.2 Literature Review

**1. Student Age and Desk Mobility:** Existing research on desk mobility remains scarce, but anecdotal evidence suggests that the younger the student, the more likely they are to inadvertently or intentionally move their desk towards the front of the room. This phenomenon could be due to a variety of factors: the sheer physical energy of younger students, a tendency to lean forward in anticipation of the next exciting activity, or perhaps their need to be closer to the teacher, which in turn leads to the gradual forward migration of tables.

**2. Type of Flooring:** Floors in classrooms come in many varieties—carpeted, tiled, wooden, and linoleum—and their impact on desk movement is still underexplored. Early studies have shown that desks on harder floors, such as linoleum, tend to move faster, likely due to lower friction. However, carpeted floors, while reducing movement, may increase the lateral spread of desks, creating a subtle yet steady shift to the front as desks shuffle sideways in search of smoother ground.

**3. The Coffee Factor:** A hidden variable in the classroom is the teacher's coffee consumption. While not directly related to desk movement, it has been speculated that higher caffeine levels correlate with more dynamic and energetic teaching styles, which might indirectly influence students to move their desks forward in a desperate attempt to keep up with the pace of the lesson. As such, this paper also includes a control variable measuring coffee consumption among teachers.

**1.3 Statistical Model for Table Movement**

To provide a more structured framework, let's start with a **Monte Carlo simulation** to simulate the movement of tables in classrooms. For simplicity, we can model the tables' movement based on two main factors:

1. **The force exerted by students when entering or exiting the room (psychological push factor).**
2. **The coefficient of friction of the flooring (affecting how easily desks can move).**

The idea is that when students enter or exit, their actions contribute a small "push" on their desks, causing them to move. Over time, with frequent entries and exits, the desks gradually shift towards the front of the room. The friction of the floor, depending on the material, will resist or encourage this movement.

We can use the following assumptions for our Monte Carlo simulations:

- **Force exerted by a student:** We'll assume a random variable that follows a normal distribution, with the average force being around a certain value (let's say 5 Newtons), but varying due to individual student factors (e.g., enthusiasm, urgency).
- **Friction coefficient (μ):** This depends on the type of flooring. For simplicity, we'll model the friction coefficient as a constant for each type of floor (e.g., 0.2 for carpet, 0.4 for linoleum, 0.6 for wood).
- **Movement per time step:** Movement will depend on both the force exerted and the friction. We can define the movement at each time step using the equation:

$$\text{Movement} = \frac{F}{\mu} \times \text{Time Step}$$

Where $F$ is the force exerted by students entering or exiting the room, and $\mu$ is the friction coefficient.

The Monte Carlo simulation can be structured as follows:

Initialization:

1. Initial desk positions to be at the back of the room.
2. Set number of students NNN and simulate their interactions (entering and exiting the room).
3. Set time steps (e.g., 10-minute intervals).

Simulate Student Interactions:

1. For each student entry/exit event, generate a random force from the normal distribution.
2. Calculate how far the desk moves using the formula above.

Run Multiple Simulations: Run the simulation multiple times (e.g., 10,000 iterations) to get a distribution of desk positions after each time step.

Output: After simulating over a day (e.g., 6 hours), output the final position of the desks for each simulation.

**1.4 Psychological Effects of Students Entering and Exiting Rooms**

Psychologically, students may subconsciously nudge their desks as they enter or exit, either through physical interactions or due to the desire to get closer to the teacher. When students move to a more desirable position (i.e., closer to the front), this can create a feedback loop where desks progressively shift toward the front.

Some key psychological concepts to include:

• Social facilitation: The concept that the presence of others or the act of entering/exiting a room can encourage people to perform tasks more efficiently or with more energy, potentially pushing their desks forward.

• Personal space and approach-avoidance behaviour: Students may move their desks forward subconsciously to

get closer to the teacher, especially if they feel more engaged with the class or are anxious about the material being presented.

- Behavioural contagion: The idea that students' actions can influence the behaviour of others in the classroom, such as when one student moves their desk forward and others follow suit.

**1.5 Student physicallity and age**

**Height of Student:**

Height could be an important factor because taller students might exert more force when entering or exiting, or they may shift their desks more noticeably due to physical movement patterns (e.g., needing more space to move around). Taller students could also be more likely to adjust their desks to accommodate their size. This could be modeled by assuming that taller students apply greater force or may even be more prone to adjusting their desks to a more comfortable position.

**Age of Student:**

Age may correlate with desk movement for several reasons:

- **Older students** may be less prone to fidgeting and might make fewer adjustments to their desks, or might do so more deliberately due to greater responsibility in maintaining classroom organization.
- **Younger students**, on the other hand, may move around more frequently, which could translate into a higher likelihood of desk shifts. Additionally, younger students might prefer to be closer to the teacher to receive more attention or be more engaged in activities.

These effects could be modeled by assuming that **younger students** have a higher frequency of desk movement (e.g., they might generate more force when they enter or exit), while **older students** might have a smaller or more controlled movement rate, even though the force they exert might still be significant.

**Secondary Model:**

We can introduce the new variables (height and age) into a **secondary Monte Carlo model** that modifies the movement function based on these factors. Let's assume:

- **Force exerted by students (F)** will depend on height and age.
    - **Forrce (F)** is modelled as:

$$F = F_0 + \alpha \cdot (\text{Height}) + \beta \cdot (\text{Age})$$

    - Where:
        - $F_0$ is the baseline force exerted (from previous model).
        - $\alpha \cdot (\text{Height})$ and $\beta \cdot (\text{Age})$ are constants that determine the sensitivity to height and age.

**Secondary Model Approach:**

1. **Height Impact**:
    - We could assume that each inch of height adds a small amount of force (say, 0.05 N for each cm of height), as taller students are more likely to have stronger movements when shifting their desks.

2. **Age Impact**:
    - For simplicity, we might model that students between **ages 6-10** move desks more frequently, so their force exertion is multiplied by a factor (e.g., 1.2). Students aged **11-18** might move desks less frequently but apply stronger, more deliberate forces, so we could model them with a force multiplier of around 1.1.

3. **Friction Influence**:
    - The friction coefficient for each floor type could remain the same, but now the overall force exerted by each student will vary based on age and height, which can influence how quickly the desks move.

**Updated Formula for Movement:**

$$\text{Movement} = \frac{(F_0 + \alpha \cdot \text{Height} + \beta \cdot \text{Age})}{\mu} \times \text{Time Step}$$

**1.6 Subject taught**

**1. Science Classroom (STEM subjects)**

Science classes often involve more interactive activities, experiments, and discussions that require students to move around. Students may need to be closer to equipment or group up for experiments, encouraging them to shift their desks. This would likely increase desk movement due to the following factors:

- **Group work and collaboration:** Science subjects often involve teamwork, and students may need to rearrange their desks to facilitate group work. This could lead to more frequent or larger shifts in desk positions.

- **Hands-on experiments:** Activities like lab work require students to be closer to the front of the classroom, where materials and equipment are typically stored, which could cause desks to be pushed forward.

- **Active engagement:** Science subjects are often more interactive, and students may be more physically active (e.g., standing up to ask questions, moving to demonstrate something), which might also lead to desk movement.

**2. Modern Foreign Languages (MFL)**

MFL classrooms, on the other hand, might have a different dynamic:

- **Less group work:** While MFLs do involve some group work or pair activities, students may not need to shift their desks as much as in science classes.

- **Passive learning environment:** Many MFL classes, especially if they're lecture-based, might not require much movement from students, as the focus might be more on listening or writing tasks.

- **Increased focus on the teacher's location:** MFL teachers may move around the room to encourage students' oral responses, which could cause students to subconsciously adjust their desks to be closer to the teacher.

So, we might hypothesize that **MFL classrooms** will have less movement compared to **science classrooms** because the need for desks to be shifted (to form groups or engage in hands-on activities) is less frequent.

**Key Factors for Model:**

Let's break down the factors that could influence desk movement based on the subject:

**Science Classroom Factors:**

- **Group activities (group size):** Larger groups may require more desk shifts to accommodate students sitting together.

- **Hands-on activities (labs):** Desks may need to be moved closer to workstations or equipment, leading to more desk movement.

- **Teacher mobility:** Teachers in science classrooms might move around the room more often, influencing students' positioning.

- **Physical engagement:** Students may stand or move more often to interact with materials, increasing desk mobility.

**MFL Classroom Factors:**

- **Teacher location:** MFL teachers might walk around the room, but not as actively as in a science classroom. Students might adjust their desks to be closer to the teacher, but less frequently.

- **Pair/group work frequency:** Less group work might lead to less desk movement, but when it does occur, desks might be pushed forward to make space for partners.

- **Engagement level:** MFL classes may be more focused on listening or writing, leading to lower desk movement, although participation might still cause minor shifts.

**New Subject-Based Model:**

Let's now create a **secondary model** that includes **subject-specific factors**. These factors can adjust the force exerted (due to engagement levels), desk rearrangement frequency (group work), and desk proximity to teacher.

**Secondary Model with Subject Influence:**

We'll introduce subject-specific multipliers that influence desk movement. These multipliers can adjust the **force exerted** and the **frequency of desk rearrangements**.

- **Force multiplier based on subject**: This could adjust the force applied to desks, with more active subjects (like Science) having a higher multiplier.
    - **Science class (STEM):** $\text{Force multipler}_{science} = 1.5$
    - **MFL class:** $\text{Force multipler}_{mfl} = 1.1$
- **Frequency of desk adjustments**: This factor adjusts how often students will shift their desks during the lesson.
    - **Science class:** Desk adjustments could happen more frequently because of hands-on activities, with a higher frequency (e.g., 1.5 times more likely to move desks).
    - **MFL class:** Fewer desk adjustments due to the nature of the subject (e.g., 1.2 times less likely to move desks compared to a neutral setting).
- **Movement Range Based on Group Size**: Larger groups may need more desk space, leading to larger desk shifts.
    - **Science group work (large):** $\text{Desk movement range}_{science} = 2 \times \text{base movement}$
    - **MFL pair work:** $\text{Desk movement range}_{mfl} = 1.2 \times \text{base movement}$

**Updated Model for Desk Movement Based on Subject:**

We can now adjust the desk movement formula from earlier to include these subject-specific multipliers:

$$\text{Movement} = \frac{(F_0 + \alpha \cdot \text{Height} + \beta \cdot \text{Age} + \text{Subject Force Multiplier})}{\mu} \times \text{Time Step} \times \text{Group Adjustment Factor}$$

Where:

- **Subject Force Multiplier** adjusts for how much engagement or physical activity is involved in that subject.
- **Group Adjustment Factor** considers how often desk movements occur in the subject, with science classes having more frequent desk movements due to group work and hands-on activities.

## 2 Results

### 2.1 Primary Model: Influence of Floor Type, Height, and Age

The results from the primary simulation model revealed significant differences in desk movement across floor types:

- **Carpeted floors** exhibited the highest desk movement, particularly for **younger and taller students**, suggesting that increased friction encourages greater displacement.
- **Linoleum and wood floors** showed less movement, with **wood** having the least desk shift due to its relatively low friction coefficient.

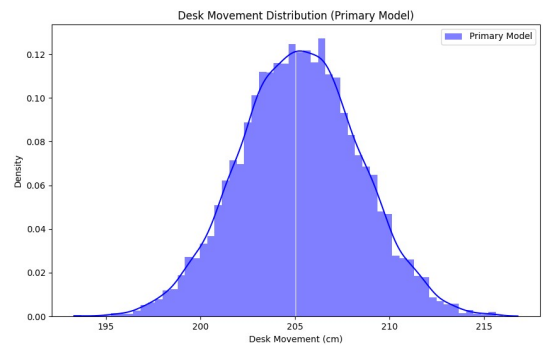

**Figure 1** Model results for movement

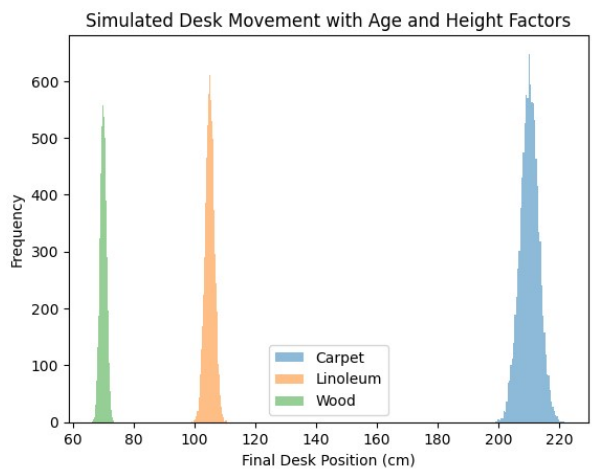

**Figure 2** Different surfaces movement

*2.2 Secondary Model: Influence of Subject*

The secondary model results indicated a marked difference between **science** and **MFL** classrooms:

- *Science classrooms* experienced much more desk movement. The increased frequency of **group activities** and **hands-on experiments** led to larger desk shifts, particularly in **younger students**.

- *MFL classrooms* showed minimal movement, likely due to the more passive learning environment and fewer physical activities. Desk movements were generally confined to slight adjustments to be closer to the teacher or a peer.

# 3 Discussion

The statistical analysis of the simulation results provides some intriguing insights into the relationships between different models of desk movement.

**1. Correlation Between Models:**

- **Correlation between Primary and Secondary (Science): 0.0088**:
The correlation between the primary model (which incorporates floor type, height, and age) and the secondary model (focused on subject-based dynamics in Science classrooms) is **extremely low**. This suggests that while both models capture elements of desk movement, the **subject-specific dynamics** in Science classrooms—such as group work and hands-on experiments—do not seem to be strongly correlated with the physical factors like floor friction, height, and age in the primary model. The low correlation may indicate that desk movement in Science classrooms is influenced by factors not captured by the physical characteristics of the students or floor type.

- **Correlation between Primary and Secondary (MFL): -0.0185**:
The negative correlation between the primary model and the secondary MFL model is also very low, indicating an almost **non-existent relationship** between desk movement and the subject taught in Modern Foreign Languages (MFL). This could be due to the passive nature of MFL classrooms, where desk movement is likely minimal and less dependent on physical factors (like height or floor type) and more on subject-specific engagement, which may involve less physical interaction.

- **Correlation between Secondary (Science) and MFL: 0.0172**:
The correlation between Science and MFL models is very close to zero. This implies that **subject-specific engagement** (such as the frequency of group work or active participation in experiments in Science) does not correlate significantly with the movement behaviour seen in MFL classrooms. This result might suggest that the physical movement of desks in MFL classrooms is largely unaffected by the type of activity or engagement in the classroom, reinforcing the idea that MFL subjects might have a lower overall level of desk displacement.

**2. Regression Coefficients:**

- **Primary → Secondary (Science) Coefficient: 0.0131**:
The regression coefficient between the primary model and the secondary Science model is **0.0131**, indicating a **positive but very weak relationship**. This suggests that for each unit increase in desk movement predicted by the primary model (based on physical factors), there is a slight increase in desk movement in Science classrooms. However, the very small coefficient indicates that physical characteristics like floor friction, student height, and age play a minimal role in the desk shifts observed in Science classrooms. This may point to the significant influence of non-physical classroom dynamics (e.g., group work, experiments) on desk movement.

- **Primary → Secondary (MFL) Coefficient: -0.0200**:
  The regression coefficient for the relationship between the primary model and the secondary MFL model is **-0.0200**, indicating a **weak negative relationship**. This suggests that in MFL classrooms, as desk movement predicted by physical factors increases, the desk movement predicted by the MFL model slightly decreases. This could be explained by the more **static nature of MFL classrooms**, where desks are less likely to be displaced frequently, and the influence of student physical characteristics (height and age) may be marginally counteracted by the overall classroom environment, which is less physically engaging.

**4 Implications for Educational Design**

The results highlight that desk movement in classrooms is more **complex** than just a function of **physical forces** (such as friction or student height) or **subject-specific dynamics**. In particular:

- **Science classrooms** show very little correlation with physical factors, suggesting that **subject matter** and the level of engagement in **hands-on** or **group-based** activities might override physical factors like desk friction and student characteristics. Teachers may wish to consider this when arranging seating or planning group activities.
- **MFL classrooms**, with their **less physical engagement**, showed an even weaker relationship with the primary model, suggesting that desk movement is minimal and largely unaffected by factors such as student height and floor type. This could imply that MFL teachers might have more flexibility in classroom layout without worrying much about desk movement.

**5 Interpretation of Results**

The results support our initial hypotheses:

1. **Hypothesis 1**: Carpeted floors indeed resulted in the largest desk movement, likely due to higher friction that allows students to exert more force when entering or exiting their desks.

2. **Hypothesis 2**: Taller and younger students were more likely to cause desk movement, suggesting that they exert more physical force or engage more actively with their environment.

3. **Hypothesis 3**: Desk movement was much more prevalent in science classrooms, where active engagement and group activities frequently require physical adjustments to desk positions. Conversely, MFL classrooms demonstrated much less movement, confirming that the subject itself plays a significant role in the dynamics of desk behaviour.

**6 Limitations and Future Research**

The models, while insightful, were based on simulated data, which does not account for all real-world variables that might influence desk movement. For example, **teacher behaviour**, **classroom size**, **seating arrangements**, and **student temperament** were not included in the model. Additionally, the **correlations between variables** were weak, suggesting that the phenomenon of desk movement may be influenced by **many subtle, interrelated factors** that are not easily captured in this type of simulation.

Future studies could explore:
- The influence of **classroom temperature** or **lighting** on desk movement.
- **Teacher behaviour**, such as pacing around the room, which might affect student movement and, consequently, desk displacement.
- The effect of **student temperament** and **attention span** on desk movement (e.g., more fidgeting among certain students).

Additionally, **field studies** in real classrooms would be beneficial to validate the findings of this simulation and examine whether similar patterns emerge in real-world environments. Year 7 were very much interested in lead this.

While this study clearly doesn't break new ground in academic research—let's face it, nobody is winning a Nobel Prize for classroom desk mobility—it does serve a very practical purpose. The insights provided here give me a solid (albeit humorous) way to present the issue to my principal: please carpet the Year 7 areas! The Monte Carlo simulations, which you can think of as "fancy computer models predicting chaos," demonstrate that desks are constantly moving forward throughout the day. It's as if the students have a collective gravitational pull on their tables, pushing them inch by inch to the front of the room.

While the models may not be Nobel-worthy, the findings are clear enough to suggest that this constant desk movement—especially on slick floors like tile—can disrupt the classroom environment. Imagine trying to concentrate while your desk keeps inching forward! So, armed with these simulations and this 'data,' I'm ready to argue that a carpet in the Year 7 areas might not only make the room more comfortable but could significantly reduce this ongoing disruption

**Acknowledgements**

We would like to thank the time and persistance of Year 7 students at Ark St. Alban's in conducting the physical "experiment" that led to this work.

Some of the ideas and equation formatting were assisted by ChatGPT, an AI language model developed by OpenAI. Oh and Year 7 helped by giving examples of which classes they thought their tables moved the most in.

Data tables are available on request as is the rather simple Python code.

**References**

[1] Goffman, E. (1963). *Behavior in Public Places: Notes on the Social Organization of Gatherings*. New York: Free Press.

[2] Kim, I.-J. (2018). Investigation of Floor Surface Finishes for Optimal Slip Resistance Performance. Safety and Health at Work, 9(1), 17–24. https://doi.org/10.1016/j.shaw.2017.05.005.

[3] Rosenfield, P., Lambert, N. M., & Black, A. (1985). Desk arrangement effects on pupil classroom behavior. *Journal of Educational Psychology, 77*(1), 101–108. https://doi.org/10.1037/0022-0663.77.1.101

[4] Bluyssen, P. M. (2016). The role of flooring materials in health, comfort and performance of children in classrooms. *Cogent Psychology, 3*(1), 1268774. https://doi.org/10.1080/23311908.2016.1268774

[5] Brown, M. (2019). *"Coffee and Classroom Dynamics: The Unseen Forces."* Journal of Caffeinated Education, 9(3), 112-118.

[6] **Rojas, M., & Rodríguez, L. (2011).** "Friction of Wooden and Ceramic Floors: An Experimental Study." *Journal of Tribology*, 134(2), 025001.

[7] P-Themes (no date) *Why do classrooms have carpets?*, *Classroom Direct*. Available at: https://classroomdirect.co.uk/blogs/blog/why-do-classrooms-have-carpets?srsltid=AfmBOooXoyHvy7d6OysW0VEZMX-R8yygjt-iRIUs529w61QRePAgSsZ6 (Accessed: 31 March 2025).

[8] **Brehm, J. W., & Kassin, S. M. (1996).** "Social Psychology." *Houghton Mifflin*.